\documentclass[referee]{svjour2}                    % onecolumn
\smartqed  % flush right qed marks, e.g. at end of proof
\usepackage{graphicx}
\usepackage{amsmath}
\usepackage{mathptmx}      % use Times fonts if available on your TeX system
\usepackage{lscape}
\usepackage{wasysym}
\journalname{Celestial Mechanics and Dynamical Astronomy}
\begin{document}

\title{Titan's rotational state}

\subtitle{The effects of a forced ``free'' resonant wobble}

\titlerunning{Titan's resonant wobble}        % if too long for running head

\author{Beno\^it Noyelles}

\institute{University of Namur, Department of Mathematics \\
Rempart de la Vierge 8, B-5000 NAMUR,BELGIUM \\
             \email{noyelles@imcce.fr} \\
           \and
              IMCCE - CNRS UMR 8028 \\
              Paris Observatory - UPMC - USTL \\
              77 avenue Denfert-Rochereau, 75014 PARIS, FRANCE
}

\date{Received: date / Accepted: date}

\maketitle

\begin{abstract}

\par In Noyelles et al. (2008, Astron. Astrophys., 478, 959-970), a resonance involving the wobble of Titan is hinted at. This paper studies this scenario and its consequences.

\par The first step is to build an accurate analytical model that would help to find the likely resonances in the rotation of every synchronous body. In this model, I take the orbital eccentricity of the body into account, as well as its variable inclination with respect to Saturn's equator. Then an analytical study using the second fundamental model of the resonance is performed to study the resonance of interest. Finally, I study the dissipative consequences of this resonance.

\par I find that this resonance may have increased the wobble of Titan by several degrees. For instance, if Titan's polar momentum $C$ is equal to $0.355MR_T^2$ ($M$ and $R_T$ being respectively Titan's mass and radius), the wobble might be forced to 41 degrees. Thanks to an original formula, I find that the dissipation associated with the forced wobble might not be negligible compared to the contribution of the eccentricity. I also suspect that, due to the forced wobble, Titan's period of rotation may be somewhat underestimated by observers. 

\par Finally, I use the analytical model presented in this paper to compute the periods of the free librations of the four Galilean satellites as well as the Saturnian satellite Rhea. For Io and Europa, the results are consistent with previous studies. For the other satellites, the periods of the free librations are respectively $186.37$ d, $23.38$ y and $30.08$ y for Ganymede, $2.44$ y, $209.32$ y and $356.54$ y for Callisto, and $51.84$ d, $2.60$ y and $3.59$ y for Rhea.

\end{abstract}

\section{Introduction}

\par As most of the major natural satellites of the Solar System planets, Titan is locked in a
spin-orbit synchronous resonance, i.e. its rotation period is very near to its orbital
period around its parent planet Saturn (see for instance Richardson et al. 2004 \cite{RLM2004}).
This corresponds to an equilibrium state known as a Cassini state. Thanks to the
Cassini fly-bys, Titan's gravity field is known well enough to study the behaviour of Titan's spin around the Cassini state. In particular, we now know Titan's oblateness coefficients $J_2$ and $C_{22}$ (Tortora et al. \cite{TAA2006}).

\par In a recent paper, Noyelles et al. (\cite{NLV2007}, hereafter cited as Paper I) give a first theory of Titan's rotation, with 3 degrees of freedom. In that work, an analytical approach and a numerical one, more complete, are compared. Moreover, some aspects of Titan's rotation are elucidated, especially a likely resonance involving the free libration of Titan's wobble.

\par In this paper, I first propose an improvement of the analytical model resulting in better agreement with the numerical results. Then I study the dynamics of the likely resonance, and discuss its implications.

\section{Convergence of the analytical study to the numerical study}

\par I firstly recall how to obtain the Hamiltonian of the problem, as explained for instance in Paper I. 3 references frames are considered: the first one $(\vec{e_1}, \vec{e_2}, \vec{e_3})$ is centered on Titan's mass barycenter and is in translation with the reference frame in which the orbital motion of Titan is described. This is a Cartesian coordinate system refering to the equatorial plane of Saturn and to the node of this plane with the ecliptic at J2000. The second frame $(\vec{n_1},\vec{n_2},\vec{n_3})$ is linked to Titan's angular momentum $\vec{G}=G\vec{n_3}$, and the third one $(\vec{f_1},\vec{f_2},\vec{f_3})$ is rigidly linked to Titan.

\par
\begin{figure}[ht]
\centering
\includegraphics[width=11.6cm]{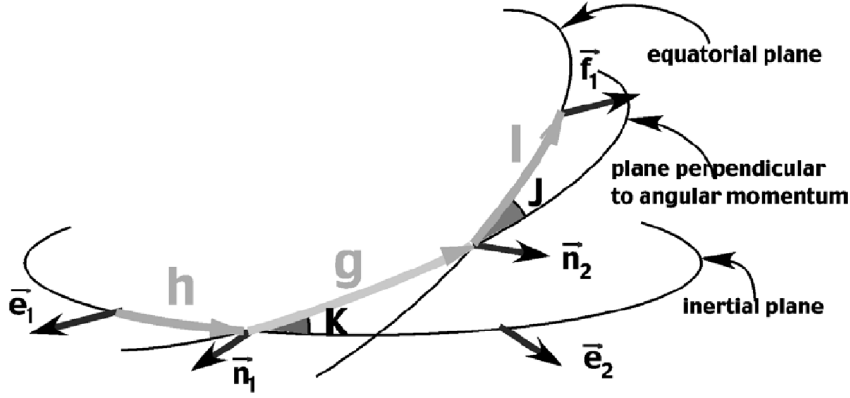}
\caption{The Andoyer variables (reproduced from Henrard \cite{Henrard2005}).}
\label{fig:angles}
\end{figure}

 The first variables that are being used are Andoyer's variables (see Andoyer 1926 \cite{Andoyer1926} and Deprit 1967 \cite{Deprit1967}), which are based on two linked sets of Euler's angles. The first set $(h,K,g)$ locates the position of the angular momentum in the first frame $(\vec{e_1},\vec{e_2},\vec{e_3})$, while the second one $(g, J, l)$ locates the body frame $(\vec{f_1},\vec{f_2},\vec{f_3})$ in the second frame $(\vec{n_1},\vec{n_2},\vec{n_3})$ (see Fig. \ref{fig:angles}).

\par The canonical set of Andoyer's variables consists of the three angular variables $l$, $g$, $h$ and their conjugated momenta defined by the norm $G$ of the angular momentum and two of its projections: 

\begin{center}
$\begin{array}{lll}
l & \hspace{2cm} & L=G\cos J \\
g & \hspace{2cm} & G \\
h & \hspace{2cm} & H=G\cos K \\
\end{array}$ \\
\end{center}
Unfortunately, these variables present two singularities: when $J=0$ (i.e. the angular momentum is colinear to $\vec{f_3}$, there is no wobble), $l$ and $g$ are undefined, and when $K=0$ (i.e. when Titan's principal axis of inertia is perpendicular to its orbital plane), $h$ and $g$ are undefined. That is why we use the modified Andoyer's variables:

\begin{equation}
\begin{array}{lll}
p=l+g+h & \hspace{2cm} & P=\frac{G}{nC} \\
r=-h & \hspace{2cm} & R=\frac{G-H}{nC}=P(1-\cos K)=2P\sin^2\frac{K}{2} \\
\xi_q=\sqrt{\frac{2Q}{nC}}\sin q & \hspace{2cm} & \eta_q=\sqrt{\frac{2Q}{nC}}\cos q \\
\end{array} \\
\label{equ:andoyer}
\end{equation}
with $Q=G-L=G(1-\cos J)=2G\sin^2\frac{J}{2}$.

\par Paper I details how to obtain the Hamiltonian of the problem in the modified Andoyer's variables. It takes account of the free rotation of the body, and of the external torque due to Saturn (see Eq.\ref{equ:hamiltonclass}).

%\par In \textbf{Paper I}, an analytical study using Henrard \& Schwanen's (2004 \cite{HS2004}) model is compared to a numerical study. These models start from the classical Hamiltonian (Eq.\ref{equ:hamiltonclass}).

\begin{equation}
  \label{equ:hamiltonclass}
\begin{split}
  \mathcal{H}=\frac{nP^2}{2}+\frac{n}{8}\big[4P-\xi_q^2-\eta_q^2\big]\Big[\frac{\gamma_1+\gamma_2}
{1-\gamma_1-\gamma_2}\xi_q^2+\frac{\gamma_1-\gamma_2}{1-\gamma_1+\gamma_2}\eta_q^2\Big] \\
+n\Big(\frac{d_0}{d}\Big)^3\bigg(1+\delta_s\Big(\frac{d_0}{d}\Big)^2\bigg)\big[\delta_1(x^2+y^2)+\delta_2(x^2-y^2)\big]
\end{split}
\end{equation}
%
%that uses the following action-angle variables :

%\begin{center}
%$\begin{array}{lll}
%p=l+g+h & \hspace{2cm} & P=\frac{G}{nC} \\
%r=-h & \hspace{2cm} & R=\frac{G-H}{nC}=P(1-\cos K)=2P\sin^2\frac{K}{2} \\
%\xi_q=\sqrt{\frac{2Q}{nC}}\sin q & \hspace{2cm} & \eta_q=\sqrt{\frac{2Q}{nC}}\cos q \\
%\end{array}$ \\
%\end{center}
%
%where $n$ is Titan's mean orbital motion, $q=-l$, and $Q=G-L=G(1-\cos J)=2G\sin^2\frac{J}{2}$.

The coefficients of the Hamiltonian are defined as follows:

\begin{center}
$\begin{array}{lll}
\gamma_1=J_2\frac{MR_T^2}{C} & \hspace{2cm} & \delta_1=-\frac{3}{2}\Big(\frac{n^*}{n}\Big)^2\gamma_1 \\
\gamma_2=2C_{22}\frac{MR_T^2}{C} & \hspace{2cm} & \delta_2=-\frac{3}{2}\Big(\frac{n^*}{n}\Big)^2\gamma_2 \\
 & \hspace{2cm} & \delta_s=\frac{5}{2}J_{2{\saturn}}\Bigg(\frac{R_{\saturn}}{d_0}\Bigg)^2  \\
\end{array}$ \\
\end{center}
where $J_{2{\saturn}}$ is Saturn's $J_2$, $d$ the distance Titan-Saturn, $d_0$ the mean value of $d$, $n^*$ the mean motion associated with $d_0$.

%\begin{figure}[ht]
%\centering
%\includegraphics[width=11.6cm]{angles.eps}
%\caption{The Andoyer variables (reproduced from Henrard \cite{Henrard2005}).}
%\label{fig:angles}
%\end{figure}

$x$ and $y$ are the first two coordinates of Saturn in the reference frame $(\vec{f_1},\vec{f_2},\vec{f_3})$ bound to Titan. If we call $x_i$, $y_i$ and $z_i$ the coordinates of Saturn's centre of mass in the reference frame $(\vec{e_1},\vec{e_2},\vec{e_3})$ defined above, we have

\begin{equation}
\left(\begin{array}{c}
x \\
y \\
z
\end{array}\right)
=R_3(-l)R_1(-J)R_3(-g)R_1(-K)R_3(-h)\left(\begin{array}{c}
x_i \\
y_i \\
z_i
\end{array}\right)
\label{equ:passage}
\end{equation}
with

\begin{equation}
R_3(\phi)=\left(\begin{array}{ccc}
\cos\phi & -\sin\phi & 0 \\
\sin\phi & \cos\phi & 0 \\
0 & 0 & 1
\end{array}\right)
\label{equ:r3}
\end{equation}
and

\begin{equation}
R_1(\phi)=\left(\begin{array}{ccc}
1 & 0 & 0 \\
0 & \cos\phi & -\sin\phi \\
0 & \sin\phi & \cos\phi
\end{array}\right).
\label{equ:r1}
\end{equation}

\par At the exact Cassini state, $\sigma=p-\lambda+\pi=0$, $\rho=r+\ascnode=0$, $\xi_q=0$ and $\eta_q=0$, $\lambda$ and $\ascnode$ being respectively Titan's mean longitude and ascending node in the frame $(\vec{e_1},\vec{e_2},\vec{e_3})$. In Paper I, Henrard \& Schwanen's model (i.e., Titan moving on a circular orbit with a constant inclination and a constant precession of the nodes) has been used to obtain the ``obliquity'' at the equilibrium $K^*$ and the three fundamental periods of the free librations around the Cassini state. Then a numerical integration has been performed over 9000 years with complete ephemeris for Titan (TASS1.6, Vienne \& Duriez 1995 \cite{VD1995})

\begin{table}[ht]
\centering
\caption{Comparison between the analytical and numerical results with $\frac{C}{MR_T^2}=0.31$, adapted from Paper I. The differences are given here relative to the numerical values, but relative to the analytical ones in Paper I.}
\begin{tabular}{l|ccc}
\hline
 & analytical & numerical & difference \\
\hline
$K^*$ (rad) & $1.1204859\times10^{-2}$ & $1.25481164\times10^{-2}$ & $10.7\%$ \\
$K^*$ (arcsec) & $2311.1681$ & $2588.2348$ & $10.7\%$ \\
$T_u$ (y) & $2.094508$ & $2.09773$ & $0.15\%$ \\
$T_v$ (y) & $167.36642$ & $167.49723$ & $0.08\%$ \\
$T_w$ (y) & $306.62399$ & $306.33602$ & $0.09\%$ \\
\hline
\end{tabular}
\label{tab:comptitNLV07}
\end{table}

\par Tab.\ref{tab:comptitNLV07}, adapted from Tab.17 in Paper I, gives a comparison between the analytical and the numerical studies, concerning the mean equilibrium position of Titan's angular momentum $K^*$, and the periods of the three free librations $u$, $v$ and $w$ around the Cassini state. The significant difference in the equilibrium position is striking, while the agreement can be considered as good for the periods of the free librations.

\par I suspect that this difference is mostly due to a too simple analytical model, that is the reason why I propose a more complete one.

\subsection{An improved model}

\par The problem is the accuracy of the model of Henrard \& Schwanen (2004 \cite{HS2004}). Henrard developed more accurate models in the particular cases of Io (2005 \cite{Henrard2005}) and Europa (2005 \cite{Henrard2005e} and \cite{Henrard2005e2}), but Henrard \& Schwanen's model is still the most accurate that could be applied to any synchronous satellite. Such a general model will not be sufficiently precise for the purposes of obtaining an ephemeris of the rotation of a body, but can be used to check the reliability of a more complete model, e.g. a numerical model like the one performed in Paper I or SONYR (Rambaux \& Bois 2004 \cite{RB2004}).

\begin{table}[ht]
  \centering
  \caption{Solution for the variable $z_6$ (eccentricity and pericentre of Titan), expressed in terms of complex exponentials (from TASS1.6, Vienne \& Duriez 1995 \cite{VD1995}).}
  \begin{tabular}{{rrrrrr}}
\hline
    n & ampl. (rad) & phase $(^{\circ})$ & frequency $(rad.y^{-1})$ & period (y) & identification \\
\hline
$1$ & $0.0289265$ & $153.988$ & $0.00893386$ & $703.30$ & $\phi_6$ \\
$2$ & $0.0001921$ & $34.663$ & $-0.00893386$ & $703.30$ & $-\phi_6$ \\
\hline
  \end{tabular}
  \label{tab:z6TASS}
\end{table}

\begin{table}[ht]
  \centering
  \caption{Solution for the variable $\zeta_6$ (inclination and ascending node of Titan), expressed in complex exponential (from TASS1.6, Vienne \& Duriez 1995 \cite{VD1995}).}
  \begin{tabular}{{rrrrrr}}
\hline
    n & ampl. (rad) & phase $(^{\circ})$ & frequency $(rad.y^{-1})$ & period (y) & identification \\
\hline
$1$ & $0.0056024$ & $184.578$ & $0.00000000$ & $\infty$ & $\Phi_0$ \\
$2$ & $0.0027899$ & $355.503$ & $-0.00893124$ & $703.51$ & $\Phi_6$ \\
$3$ & $0.0001312$ & $289.015$ & $-0.00192554$ & $3263.07$ & $\Phi_8$ \\
\hline
  \end{tabular}
  \label{tab:zeta6TASS}
\end{table}

\par Tab. \ref{tab:z6TASS} and \ref{tab:zeta6TASS} come from TASS1.6 ephemeris and give the main terms of the solutions for $z_6=e\exp(\sqrt{-1}\varpi)$ (eccentricity and pericentre of Titan) and $\zeta_6=\Gamma\exp(\sqrt{-1}\ascnode)$ (inclination and ascending node of Titan) with $\Gamma=\sin\frac{I}{2}$. We can see that the second term in $z_6$ has a very small amplitude compared to the first one. On the contrary, the second term in $\zeta_6$ cannot be neglected compared to the first term, because its amplitude is half the amplitude of the first term. The amplitude of the third term is much smaller.

\par For this reason I choose to write $z_6$ and $\zeta_6$ as follows:

\begin{equation}
  \label{eq:z6simp}
  z_6=e_1\exp(\sqrt{-1}\phi_6)
\end{equation}

\begin{equation}
  \label{eq:zeta6simp}
  \zeta_6=\Gamma_0\exp(\sqrt{-1}\Phi_0)+\Gamma_1\exp(\sqrt{-1}\Phi_6)
\end{equation}
with $e_1=0.0289265$, $\Gamma_0=0.0056024$ and $\Gamma_1=0.0027899$. The analytical study performed in Paper I used $e_1=0$ and $\Gamma_1=0$, as in Henrard \& Schwanen's work (2004 \cite{HS2004}).

\par With this model, the way to proceed is the same as in the other studies (Henrard 2005 \cite{Henrard2005} and \cite{Henrard2005e}). The equations have been developed with Maple software to the third degree in inclination and second in eccentricity. The Cassini state corresponds to the equilibrium of the Hamiltonian : 

\begin{equation}
\begin{split}
\mathcal{H}=\frac{nP^2}{2}-nP+\dot{\ascnode}R+n\delta_1(1+\delta_s)[a_1\sin^2K+a_2\sin K\cos K\cos\rho \\
+a_3\cos2\rho(1-\cos2K)]+n\delta_2(1+\delta_s)[b_1(1+\cos K)^2\cos2\sigma \\
+b_2\sin K(1+\cos K)\cos(2\sigma+\rho)+b_3\sin^2K\cos(2\sigma+2\rho) \\
+b_4\sin K(1-\cos K)\cos(2\sigma+3\rho)+b_5(1-\cos K)^2\cos(2\sigma+4\rho)]
\end{split}
\label{equ:hamsimp}
\end{equation}

with $\sigma=0$, $\rho=0$, $\xi_q=0$ and $\eta_q=0$, the coefficients $a_i$, $b_i$ being now:

\begin{equation}
  \label{eq:a1}
  a_1=-\frac{1}{2}+3\big(\Gamma_0^2+\Gamma_1^2\big)-\frac{3}{4}e_1^2
\end{equation}

\begin{equation}
  \label{eq:a2}
  a_2=2\Gamma_0+\frac{\Gamma_1^2}{2\Gamma_0}-\frac{45}{4}\Gamma_0\Gamma_1^2+3\Gamma_0e_1^2+\frac{3\Gamma_1^2e_1^2}{4\Gamma_0}-5\Gamma_0^3
\end{equation}

\begin{equation}
  \label{eq:a3}
  a_3=\frac{\Gamma_0^2+\Gamma_1^2}{2}
\end{equation}

\begin{equation}
  \label{eq:b1}
  b_1=\frac{1}{4}-\frac{\Gamma_0^2+\Gamma_1^2}{2}-\frac{5}{8}e_1^2
\end{equation}

\begin{equation}
  \label{eq:b2}
  b_2=\Gamma_0+\frac{\Gamma_1^2}{4\Gamma_0}-\frac{5}{2}\Gamma_0e_1^2-\frac{27}{8}\Gamma_0\Gamma_1^2-\frac{5\Gamma_1^2e_1^2}{8\Gamma_0}-\frac{3}{2}\Gamma_0^3
\end{equation}

\begin{equation}
  \label{eq:b3}
  b_3=\frac{3}{2}\big(\Gamma_0^2+\Gamma_1^2\big)
\end{equation}

\begin{equation}
  \label{eq:b4}
  b_4=\Gamma_0^3+\frac{9}{4}\Gamma_0\Gamma_1^2
\end{equation}

\par $b_5$ does not appear because of the truncation to the third degree in inclination, so I propose to use the expression given in Paper I, i.e. :

\begin{equation}
  \label{eq:b5}
  b_5=\frac{\Gamma_0^4}{4}
\end{equation}
Then, the three fundamental periods of libration around the exact equilibrium should be processed the same way as in \cite{HS2004}, \cite{Henrard2005}, \cite{Henrard2005e} and Paper I.

\subsection{Results}

\par I here present the numerical applications of the model, with the initial parameters given in Tab.\ref{tab:data}.

\begin{table}[ht]
\centering
\caption{Physical and dynamical parameters. They are the same as in Paper I, except for $e_1$ and $\Gamma_1$ that have been added in this paper and come from TASS1.6 (Vienne \& Duriez 1995 \cite{VD1995}). The chosen value of $\frac{C}{MR_T^2}$ is arbitrary. The assumption of the hydrostatic equilibrium is up to now doubtful for Titan, in particular because it would imply $J_2=\frac{10}{3}C_{22}$, and so cannot be used for estimating $\frac{C}{MR_T^2}$. I have chosen this value firstly because it is not physically absurd, and secondly because it gives a dynamical system far from the resonance hinted in Paper I. Later in the paper I change this value, to be able to study the resonance itself.}
\begin{tabular}{l|ll}
\hline
$n$ & $143.9240478491399 rad.y^{-1}$ & TASS1.6  \cite{VD1995} \\
$e_1$ & $0.0289265$ & TASS1.6 \cite{VD1995} \\
$\Gamma_0$ & $5.6024\times10^{-3}$ & TASS1.6 \cite{VD1995} \\
$\Gamma_1$ & $2.7899\times10^{-3}$ & TASS1.6 \cite{VD1995} \\
$R_{\saturn}$ & $58232$ km & IAU 2000 \cite{SAB2002} \\
$J_{2\saturn}$ & $1.6298\times10^{-2}$ & Pioneer \& Voyager \cite{CA1989} \\
$M$ & $2.36638\times10^{-4}M_{\saturn}$ & Pioneer \& Voyager \cite{CA1989} \\
$R_T$ & $2575$ km & IAU 2000 \cite{SAB2002} \\
$\mathcal{G}M_{\saturn}$ & $3.77747586645\times10^{22}.km^3.y^{-2}$ & Pioneer, Voyager + IERS 2003 \\
$J_{2}$ & $(3.15\pm0.32)\times10^{-5}$ & Cassini \cite{TAA2006} \\
$C_{22}$ & $(1.1235\pm0.0061)\times10^{-5}$ & Cassini \cite{TAA2006} \\
$\frac{C}{MR_T^2}$ & $0.31$ &  \\
\hline
\end{tabular}
\label{tab:data}
\end{table}

\par Tab. \ref{tab:comptousmod} gives a comparison between the circular model (originally given by Henrard \& Schwanen 2004 \cite{HS2004}), the numerical model from Paper I and the model given in this paper, with or without $e_1$, and with or without $\Gamma_1$.

%\begin{table}[ht]
%  \centering
%  \caption{Comparison between the different models, with $\frac{C}{MR_T^2}=0.31$. The relative difference with the results coming from the numerical model are given into parenthesis.}
%  \begin{tabular}{l|c}
%    \hline
%Model & $K^*$ (rad) \\
%\hline
%Circular \cite{HS2004} & $1.1204859\times10^{-2}$ $(10.7\%)$ \\
%With $e_1=0$ & $1.1899571\times10^{-2}$ $(5.17\%)$ \\
%With $\Gamma_1=0$ & $1.1204858\times10^{-2}$ $(10.7\%)$ \\
%$e_1$,$\Gamma_1 \neq 0$ & $1.1899570\times10^{-2}$ $(5.17\%)$ \\
%Numerical \cite{NLV2007} & $1.2548116\times10^{-2}$ \\
%  \end{tabular}
%  \begin{tabular}{l|ccc}
%    \hline
%Model & $T_u$ $(y)$ & $T_v$ $(y)$ & $T_w$ $(y)$ \\
%\hline
%Circular \cite{HS2004} & $2.09451$ $(0.15\%)$ & $167.36642$ $(0.08\%)$ & $306.62399$ $(0.09\%)$ \\
%With $e_1=0$ & $2.09452$ $(0.15\%)$ & $167.36955$ $(0.08\%)$ & $306.64198$ $(0.1\%)$ \\
%With $\Gamma_1=0$ & $2.09670$ $(0.05\%)$ & $167.38996$ $(0.06\%)$ & $306.64204$ $(0.1\%)$ \\
%$e_1$,$\Gamma_1\neq 0$ & $2.09671$ $(0.05\%)$ & $167.39269$ $(0.06\%)$ & $306.64198$ $(0.1\%)$ \\
%Numerical \cite{NLV2007} & $2.09773$ & $167.49723$ & $306.33602$ \\
%  \end{tabular}
%  \label{tab:comptousmod}
%\end{table}

\begin{landscape}
\begin{table}[ht]
  \centering
  \caption{Comparison between the different models, with $\frac{C}{MR_T^2}=0.31$. The relative differences with the results from the numerical model are given in parentheses.}
%  \begin{tabular}{l|c}
%    \hline
%Model & $K^*$ (rad) \\
%\hline
%Circular \cite{HS2004} & $1.1204859\times10^{-2}$ $(10.7\%)$ \\
%With $e_1=0$ & $1.1899571\times10^{-2}$ $(5.17\%)$ \\
%With $\Gamma_1=0$ & $1.1204858\times10^{-2}$ $(10.7\%)$ \\
%$e_1$,$\Gamma_1 \neq 0$ & $1.1899570\times10^{-2}$ $(5.17\%)$ \\
%Numerical \cite{NLV2007} & $1.2548116\times10^{-2}$ \\
%  \end{tabular}
  \begin{tabular}{l|cccc}
    \hline
Model &  $K^*$ (arcsec)  & $T_u$ $(y)$ & $T_v$ $(y)$ & $T_w$ $(y)$ \\
\hline
Circular \cite{HS2004}& $2311.1681$ $(10.7\%)$ & $2.09451$ $(0.15\%)$ & $167.36642$ $(0.08\%)$ & $306.62399$ $(0.09\%)$ \\
With $e_1=0$  & $2454.4627$ $(5.17\%)$ & $2.09452$ $(0.15\%)$ & $167.36955$ $(0.08\%)$ & $306.64198$ $(0.1\%)$ \\
With $\Gamma_1=0$  & $2311.1679$ $(10.7\%)$ & $2.09670$ $(0.05\%)$ & $167.38996$ $(0.06\%)$ & $306.64204$ $(0.1\%)$ \\
$e_1$,$\Gamma_1\neq 0$  & $2454.4625$ $(5.17\%)$ & $2.09671$ $(0.05\%)$ & $167.39269$ $(0.06\%)$ & $306.64198$ $(0.1\%)$ \\
Numerical \cite{NLV2007} & $2588.2348$ & $2.09773$ & $167.49723$ & $306.33602$ \\
  \end{tabular}
  \label{tab:comptousmod}
\end{table}
\end{landscape}

\par We can see that the values of the equilibrium ``obliquity'' $K^*$ and of the first fundamental period $T_u$ are significantly improved. In particular, taking $\Gamma_1$ into account helps to approximate $K^*$, while taking $e_1$ into account improves the determination of $T_u$.

\subsection{Expressing the free solution}

\par In this part I use the model described in this paper to explain the main free terms obtained numerically in Paper I.

\par The main terms of the free solution come from the Hamiltonian :

\begin{equation}
  \label{eq:hamfrek}
  \mathcal{N}=\omega_uU+\omega_vV+\omega_wW
\end{equation}
its derivation being explained in Paper I. Using the different canonical transformations that have been used between the variables $(\sigma,\rho,\xi_q,P,R,\eta_q)$ and \\
$(u,v,w,U,V,W)$, we have:

\begin{equation}
  \label{eq:equu}
  u(t)=\omega_ut+u_0, U(t)=U_0
\end{equation}

\begin{equation}
  \label{eq:equv}
  v(t)=\omega_vt+v_0, V(t)=V_0
\end{equation}

\begin{equation}
  \label{eq:equw}
  w(t)=\omega_wt+w_0, W(t)=W_0
\end{equation}
then

\begin{equation}
  \label{eq:xy1}
  x_1=\sqrt{2U_0U^*}\sin(\omega_ut+u_0), y_1=\sqrt{\frac{2U_0}{U^*}}\cos(\omega_ut+u_0)
\end{equation}

\begin{equation}
  \label{eq:xy2}
  x_2=\sqrt{2V_0V^*}\sin(\omega_vt+v_0), y_2=\sqrt{\frac{2V_0}{V^*}}\cos(\omega_vt+v_0)
\end{equation}

\begin{equation}
  \label{eq:xy3}
  x_3=\sqrt{2W_0W^*}\sin(\omega_wt+w_0), y_3=\sqrt{\frac{2W_0}{W^*}}\cos(\omega_wt+w_0)
\end{equation}
and finally

\begin{equation}
  \label{eq:sigma}
  \sigma(t)=\sqrt{2U_0U^*}\sin(\omega_ut+u_0)-\beta\sqrt{2V_0V^*}\sin(\omega_vt+v_0)
\end{equation}

\begin{equation}
  \label{eq:rho}
  \rho(t)=\alpha\sqrt{2U_0U^*}\sin(\omega_ut+u_0)+(1-\alpha\beta)\sqrt{2V_0V^*}\sin(\omega_vt+v_0)
\end{equation}

\begin{equation}
  \label{eq:eta}
  \eta_q(t)=\sqrt{\frac{2W_0}{W^*}}\cos(\omega_wt+w_0)
\end{equation}

\begin{equation}
  \label{eq:xi}
  \xi_q(t)=\sqrt{2W_0W^*}\sin(\omega_wt+w_0)
\end{equation}

\begin{equation}
  \label{eq:P}
  P(t)=P^*+(1-\alpha\beta)\sqrt{\frac{2U_0}{U^*}}\cos(\omega_ut+u_0)-\alpha\sqrt{\frac{2V_0}{V^*}}\cos(\omega_vt+v_0)
\end{equation}

\begin{equation}
  \label{eq:R}
  R(t)=R^*+\beta\sqrt{\frac{2U_0}{U^*}}\cos(\omega_ut+u_0)+\sqrt{\frac{2V_0}{V^*}}\cos(\omega_vt+v_0)
\end{equation}
where $U_0$, $V_0$ and $W_0$ are the constant real amplitudes associated with the 3 fundamental proper modes, and $u_0$, $v_0$ and $w_0$ are the phases at the time origin (i.e. J1980, JD 2444240). The constants $\alpha$, $\beta$, $U^*$, $V^*$ and $W^*$ are used in the changes of variables and are defined in Paper I, and $R^*$ and $P^*$ are respectively the equilibrium values of $R$ and $P$.

\par Comparing Eq.(\ref{eq:sigma})-(\ref{eq:R}) to the numerical solutions given in Paper I gives the following equations:
from $P$:

\begin{equation}
  \label{eq:equa1}
  (1-\alpha\beta)\sqrt{\frac{2U_0}{U^*}}=9.9514\times10^{-5}
\end{equation}
from $R$:

\begin{equation}
  \label{eq:equa2}
  \sqrt{\frac{2V_0}{V^*}}=2.28952572\times10^{-5}
\end{equation}
from $\eta_q$:

\begin{equation}
  \label{eq:equa3}
  \sqrt{\frac{2W_0}{W^*}}=1.514080315\times10^{-3}
\end{equation}
from $\xi_q$:

\begin{equation}
  \label{eq:equa4}
  \sqrt{2W_0W^*}=3.10703141\times10^{-4}
\end{equation}
from $\sigma$:

\begin{equation}
  \label{eq:equa5}
  \sqrt{2U_0U^*}=4.78176461\times10^{-3}
\end{equation}

\begin{equation}
  \label{eq:equa6}
  |\beta|\sqrt{2V_0V^*}=1.147635\times10^{-5}
\end{equation}
and from $\rho$:

\begin{equation}
  \label{eq:equa7}
  (1-\alpha\beta)\sqrt{2V_0V^*}=0.18089837
\end{equation}

\begin{table}[h]
  \centering
  \caption{Numerical values of the useful parameters, computed with the model presented in this paper. They are dimensionless parameters.}
  \begin{tabular}{|l|l|}
    \hline
$\alpha$ & $-7.6905124339\times10^{-9}$  \\
\hline
$\beta$ & $7.0794875675\times10^{-5}$ \\
\hline
$U^*$ & $48.03220934680$  \\
\hline
$V^*$ & $7062.426811769$ \\
\hline
$W^*$ & $0.2046085670251$ \\
\hline
  \end{tabular}
  \label{tab:lesparam}
\end{table}

\par Using Eq.\ref{eq:equa1}-\ref{eq:equa7} and the numerical parameters given in Tab.\ref{tab:lesparam}, I computed the values of the amplitudes of the free librations $U_0$, $V_0$ and $W_0$ presented in Paper I for $\frac{C}{MR_T^2}=0.31$. The results are given in Tab.\ref{tab:resul}.

\begin{table}[h]
  \centering
  \caption{Numerical determinations of $U_0$, $V_0$ and $W_0$.}
  \begin{tabular}{|l|l|l|}
    \hline
Equation & amplitude (dimensionless) & phase \\
\hline
\ref{eq:equa1} & $U_0=2.378323538673\times10^{-7}$ & $u_0=63.00^{\circ}$ \\
\ref{eq:equa5} & $U_0=2.380202066113\times10^{-7}$ & $u_0=63.00^{\circ}$ \\
\hline
\ref{eq:equa2} & $V_0=1.851036650588\times10^{-6}$ & $v_0=174.72^{\circ}$ \\
\ref{eq:equa6} & $V_0=1.860458412872\times10^{-6}$ & $v_0=174.40^{\circ}$ \\
\ref{eq:equa7} & $V_0=2.316782965743\times10^{-6}$ & $v_0=175.64^{\circ}$ \\
\hline
\ref{eq:equa3} & $W_0=2.345263498798\times10^{-7}$ & $w_0=-51.69^{\circ}$ \\
\ref{eq:equa4} & $W_0=2.359051803911\times10^{-7}$ & $w_0=-51.69^{\circ}$ \\
\hline
  \end{tabular}
  \label{tab:resul}
\end{table}

\par We can see a very good agreement between the different determinations of $U_0$, $V_0$ and $W_0$, which tends to confirm the agreement between the analytical and the numerical study. However, Eq.\ref{eq:equa7} gives a slightly different result from Eq.\ref{eq:equa2} and \ref{eq:equa6}, both for $V_0$ and $v_0$. This might be due to a contribution in $\rho$ whose period is very near $T_v$ and so cannot be separated from $v$ over only 9000 years. We are limited to this interval of time because it is the limit of validity of the TASS1.6 ephemeris, as stated by Vienne \& Duriez (1995 \cite{VD1995}).

\section{The resonance}

\par As seen in Paper I, Titan's rotation seems to encounter a quasi-resonant state when $T_w$, i.e. the period of the free libration associated with the wobble $J$, is near 350 years. This occurs when $\frac{C}{MR_T^2} \approx 0.35$. This part aims at first identifying the resonant argument, then the associated Hamiltonian, and finally to study the associated dynamics.

\subsection{Identification of the resonant argument}

\par
\begin{figure}[ht]
  \centering
  \begin{tabular}{cc}
    \includegraphics[width=5.4cm]{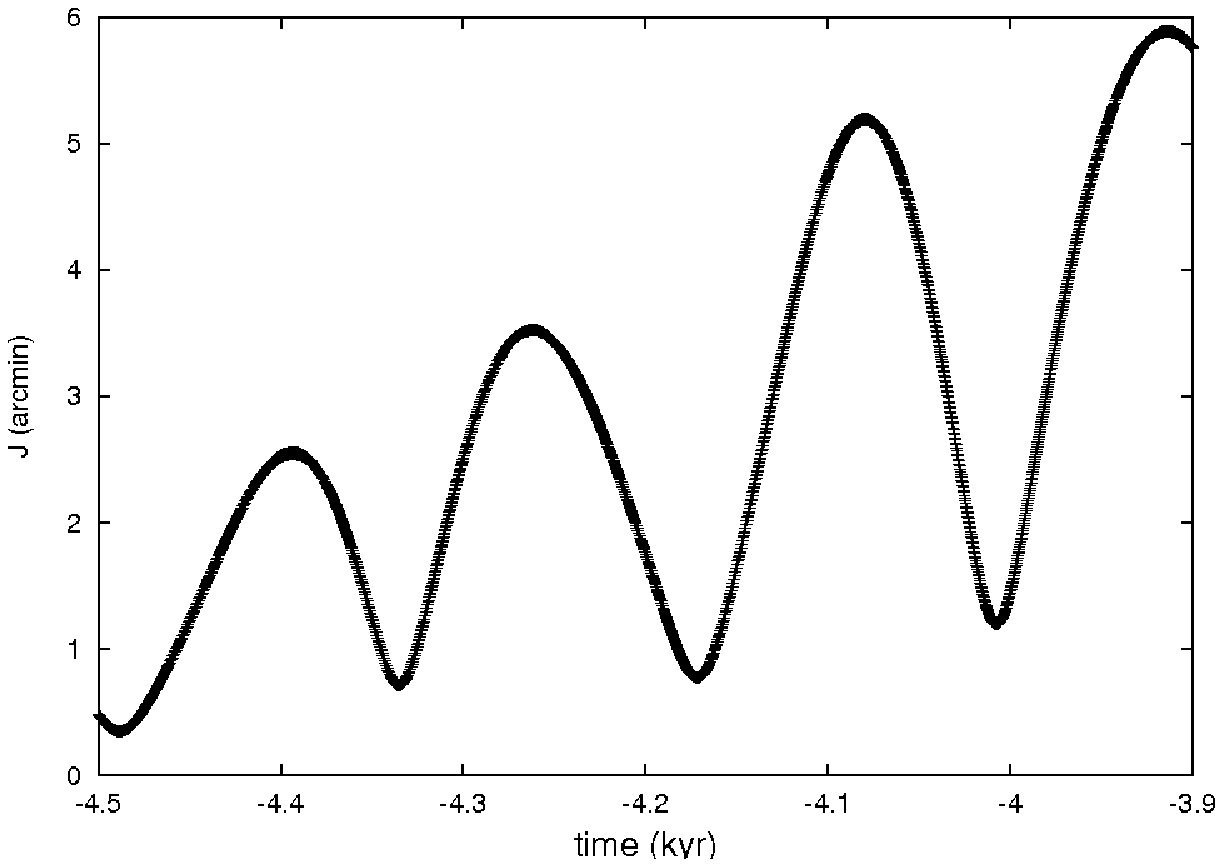} & \includegraphics[width=5.4cm]{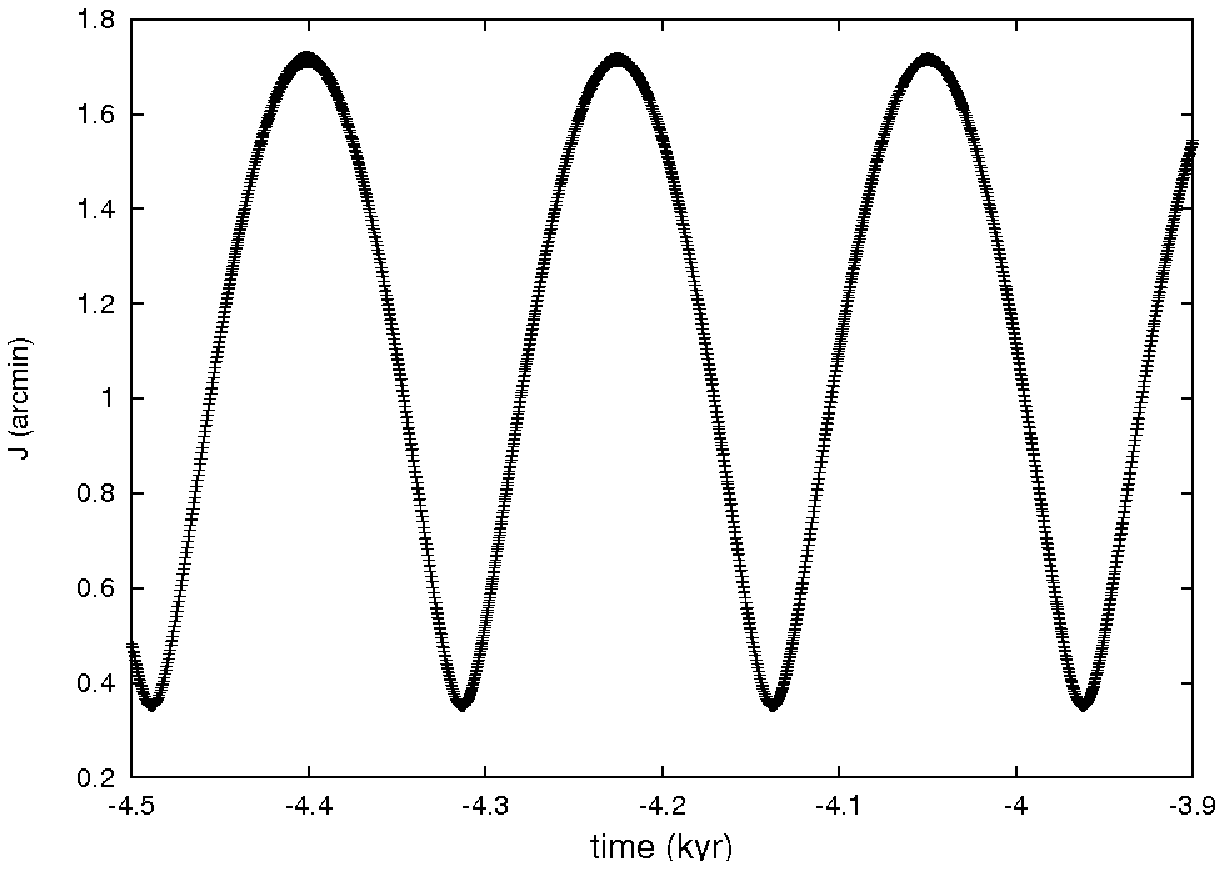} \\
    (a) TASS1.6 & (b) circular orbit $(e_1=0, \Gamma_1=0)$ \\
    \includegraphics[width=5.4cm]{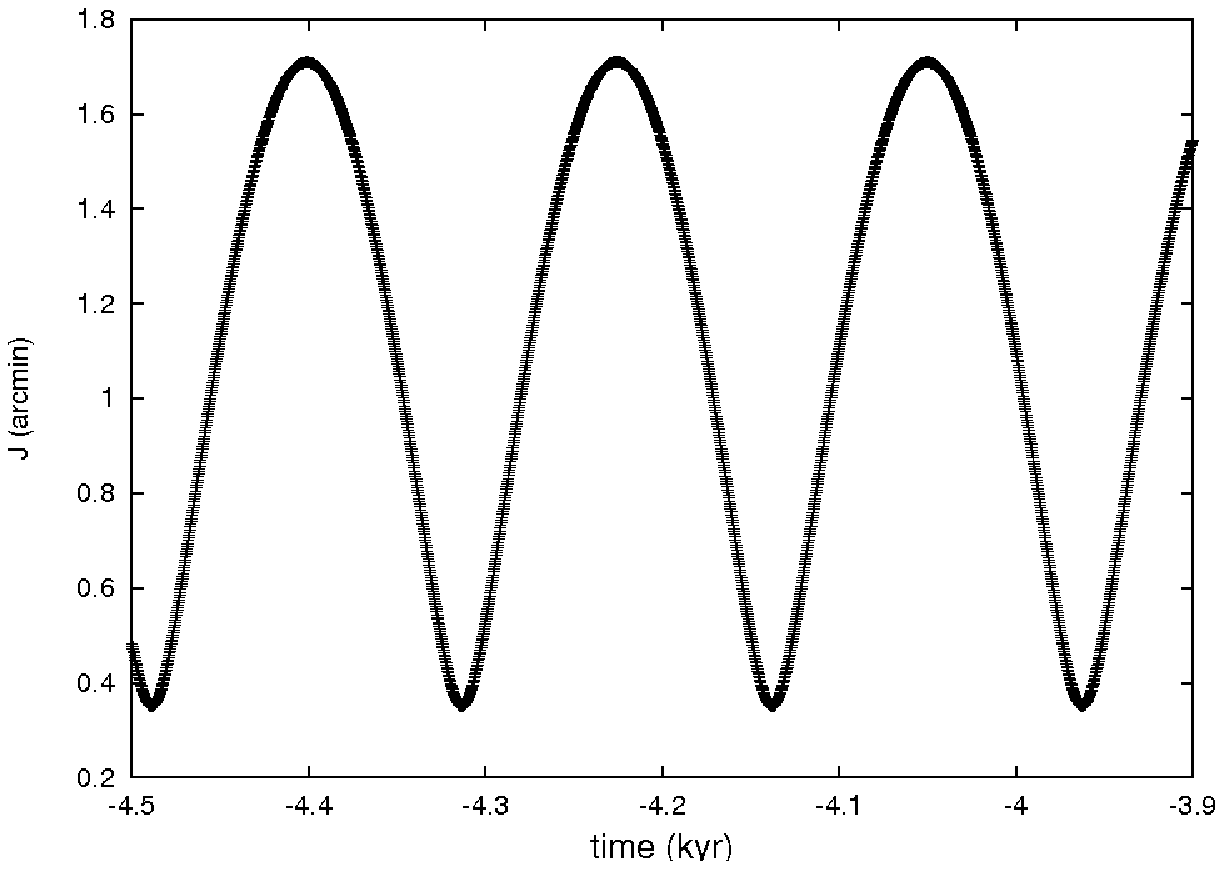} & \includegraphics[width=5.4cm]{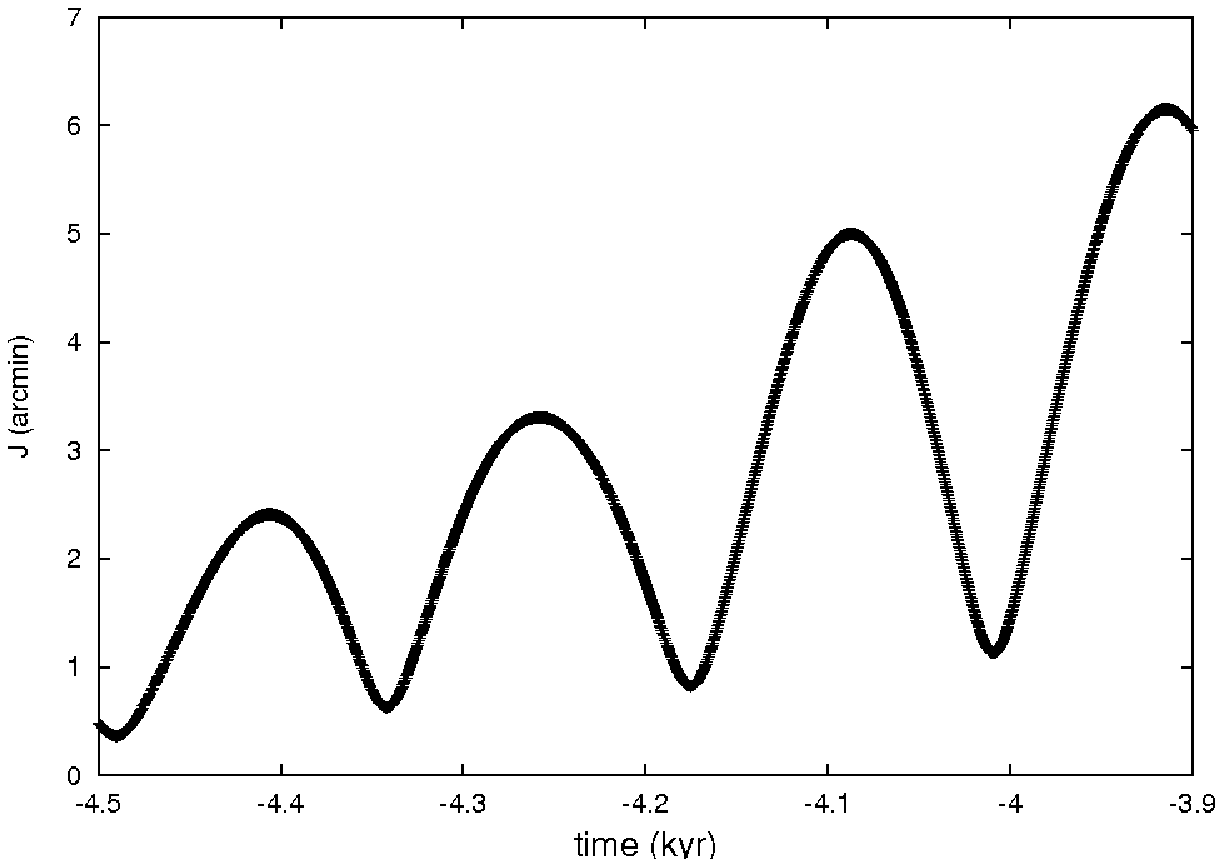} \\
    (c) $e_1=0$, $\Gamma_1 \ne 0$ & (d) $e_1 \ne 0$, $\Gamma_1 \ne 0$ \\
  \end{tabular}
  \caption{Quasi-resonant behaviour of Titan's wobble $J$, with $\frac{C}{MR_T^2}=0.355$ and different models of Titan's orbital motion.}
  \label{fig:Jresonne}
\end{figure}

 From the quasiperiodic decomposition of $\eta_q+\sqrt{-1}\xi_q$, a periodic contribution
whose period is about $351.7$ years is likely to lock the system in a resonance with $w$.
Unfortunately, such a period might result from different integer combinations of proper modes of
TASS1.6, more precisely this might be the period of $-2\Phi_6$, $2\phi_6$ or $\phi_6-\Phi_6$. In
TASS1.6, the amplitudes associated with $\phi_6$ and $\Phi_6$ are respectively $e_1$
and $\Gamma_1$. So, we tried several numerical computations with/without $e_1$, with/without
$\Gamma_1$, to check when the quasi-resonant behaviour occurs (see Fig.\ref{fig:Jresonne}). We used $\frac{C}{MR_T^2}=0.355$, to be closer to the exact resonance.

\par These numerical integrations have been performed over 600 years, which is sufficient visually to discriminate quasi-resonant behaviour from ``normal'' behaviour. It appears clearly that the quasi-resonant behaviour is reproduced with the model presented in this paper, i.e. with one term in eccentricity and two in inclination. So, the argument inducing the quasi-resonant state should be $f=w+\Phi_6-\phi_6$.

\subsection{The Hamiltonian}

\par I here intend to express the Hamiltonian of the problem, considering the resonant argument (or quasi-resonant) as a slow argument that should not be averaged. 

\par I start from the following Hamiltonian :

\begin{equation}
  \label{eq:hamilkp}
  \mathcal{K}=\mathcal{N}+\mathcal{P}=\omega_uU+\omega_vV+\omega_wW+\mathcal{P}
\end{equation}
where $\mathcal{K}$ is the Hamiltonian of the complete system, $\mathcal{N}$ the Hamiltonian of the free solution to the first order expressed in (Eq.\ref{eq:hamfrek}), and $\mathcal{P}$ is the remaining part of $\mathcal{K}$. So, the Hamiltonian is centered on the exact Cassini state and the canonical variables are $(u,v,w,U,V,W)$. In order for the resonant argument $f$ to appear explicitly, I use the following new set of canonical variables :

\begin{center}
$\begin{array}{lll}
u & \hspace{2cm} & U \\
v & \hspace{2cm} & V \\
\theta=w+\Phi_6-\phi_6=f & \hspace{2cm} & \Theta=W \\
\end{array}$ \\
\end{center}

\par Since this change of variables depends explicitly on time, I should add the quantity $\big(\dot{\Phi}_6-\dot{\phi}_6\big)\Theta$ to the new Hamiltonian $\mathcal{T}$, which then becomes : 

\begin{equation}
  \label{eq:T1}
  \mathcal{T}=\omega_uU+\omega_vV+\big(\omega_w+\dot{\Phi}_6-\dot{\phi}_6\big)\Theta+\mathcal{T}_2
\end{equation}

\par Considering that $f=\theta$ is the only slow argument, every periodic term (except of course the slow one) averages to a constant value and is dropped from the Hamiltonian, so that $\mathcal{T}$ becomes:

\begin{equation}
  \label{eq:T2}
  \mathcal{T}=\psi \Theta+\mu \Theta^2+\epsilon\sqrt{2\Theta}\cos \theta
\end{equation}
The expressions for $\psi$, $\mu$ and $\epsilon$ are given in the Appendix.

\par This Hamiltonian is the Second Fundamental Model of Resonance of Henrard and Lema\^itre (1983 \cite{HL1983}) and we can now study its dynamical consequences.

\subsection{Results}

\par According to (Henrard and Lema\^itre 1983 \cite{HL1983}), the resonance associated with the Hamiltonian (\ref{eq:T2}) has two stable equilibria if a critical parameter, $\delta=-1-sign(\psi\mu)\Big|\frac{4}{27}\frac{\psi^3}{\mu\epsilon^2}\Big|^{\frac{1}{3}}$, is positive. These equilibria correspond to the roots of the cubic equation $x^3-3(\delta+1)x-2=0$ with $x=\sqrt{2R}$ and $R=\Big|\frac{2\mu}{\epsilon}\Big|^{\frac{2}{3}}\Theta$. Only one of these roots is positive, it is of course the one that is interesting and it corresponds to a stable equilibrium.

\begin{table}[ht]
  \centering
  \caption{Values of $\delta$ and of the forced amplitude of the libration with different values of $\frac{C}{MR_T^2}$. The values of the other parameters are given in Tab.\ref{tab:data}. The mean wobble $<J>$ has been computed using Eq.\ref{eq:meanJ}.}
  \begin{tabular}{l|c|c|c}
    $\frac{C}{MR_T^2}$ & $\delta$ & $W_0$ (forced) & $<J>$ \\
\hline
$0.34$ & $5349.4$ & (no real solution) & \\
$0.35$ & $1909.6$ & $0.342$ & $80.368^{\circ}$ \\
$0.355$ & $189.69$ & $0.108$ & $40.702^{\circ}$ \\
$0.3555$ & $17.705$ & $0.034$ & $22.337^{\circ}$ \\
$0.355551$ & $0.1616$ & $0.010$ & $12.034^{\circ}$ \\
$0.35555146967191$ & $1.6\times10^{-10}$ & $0.009$ & $11.413^{\circ}$ \\
$0.35555146967192$ & $-3.3\times10^{-9}$ & (no resonance) & \\
  \end{tabular}
  \label{tab:lesdelta}
\end{table}

\par Different values of $\delta$ and of the forced amplitude of the wobble are given in Tab.\ref{tab:lesdelta}. We see that, under the resonance, the wobble $J$ is increasing if $\frac{C}{MR_T^2}$ is decreasing.

\par The mean wobble $<J>$ is obtained as follows: we have $\xi_q^2+\eta_q^2=\frac{4G}{nC}\sin^2\frac{J}{2}$ from Eq.\ref{equ:andoyer}. Assuming that $G \approx nC$, we get $\xi_q^2+\eta_q^2=4\sin^2\frac{J}{2}$. Moreover, it comes from Eq.\ref{eq:eta}-\ref{eq:xi} that $\xi_q^2+\eta_q^2=\frac{2W_0}{W^*}\big(\cos^2(\omega_wt+w_0)+W^{*2}\sin^2(\omega_wt+w_0)\big)$. Equating these two last relations we get \\
$J=2arcsin\Big(\sqrt{\frac{W_0}{2W^*}}\sqrt{\cos^2(\omega_wt+w_0)+W^{*2}\sin^2(\omega_wt+w_0)}\Big).$ Finally $<J>$ is obtained in averaging this last relation, i.e.

\begin{equation}
\label{eq:meanJ}
<J>=\frac{1}{2\pi}\int_{0}^{2\pi}2arcsin\Big(\sqrt{\frac{W_0}{2W^*}}\sqrt{\cos^2(\tau)+W^{*2}\sin^2(\tau)}\Big)d\tau
\end{equation}

\par Since it is a first-order resonance, it is a strong resonance and the capture into it is highly likely if the ratio $\frac{C}{MR_T^2}$ actually has the values given in Tab.\ref{tab:lesdelta}, which is still unknown.

%\par These computations have been performed \textbf{treating $e_1$ and $\Gamma_1$ as constant}. If the resonance actually \textbf{occurred} (what should be known soon thanks to \textbf{the} Cassini mission), it would be interesting to study this resonance considering $e_1$ and $\Gamma_1$ as varying actions. It cannot be excluded that they are in fact forced by the resonance. This could explain what the orbit of Titan is currently far \textbf{from} circular.

\section{Discussion}

\subsection{Consequences for Titan's internal dissipation}

\par Here I estimate the effect of a forced wobble on Titan's internal dissipation. I use for that Wisdom's work (2004 \cite{Wisdom2004}), in which the reader can find a more detailed explanation, as well as in (Murray and Dermott 1999 \cite{MD1999}).

\par Time-dependent tidal distortion of a body leads to internal heating. Let $U_T$ be the tide-raising potential. The total dissipated energy is given by integration over the whole volume of the satellite (here assumed to be Titan), i.e.
\begin{equation}
  \frac{dE}{dt}=-\int_{body}\rho\vec{v}.\vec{\nabla U_T}dV
  \label{eq:Wisdom1}
\end{equation}
where $\rho$ is the density, $dV$ the volume element, and $\vec{v}$ its velocity. Assuming that the body is incompressible (it is a good approximation, see Peale and Cassen 1978 \cite{PC1978}) and that the density $\rho$ is constant, Eq.\ref{eq:Wisdom1} becomes:

\begin{equation}
  \frac{dE}{dt}=-\frac{\rho h_s}{g}\int_{surface}U_T\frac{d }{dt}(U'_T)dS
  \label{eq:Wisdom2}
\end{equation}
where $U_T$ is the tide-raising gravity gradient potential, $U'_T$ is the same potential including a time delay $\Delta$ due to the tidal response lags of Titan, $h_s$ is Titan's displacement Love number, and $g$ the local acceleration of gravity.

\par The tide-raising gravity potential is

\begin{equation}
  \label{eq:utez}
  U_T=-\frac{GM_{\saturn}R^2}{r^3}P_2(\cos\alpha)
\end{equation}
where $P_2$ is the second Legendre polynomial ($P_2(x)=\frac{3}{2}x^2-\frac{1}{2}$), $\alpha$ is the angle at the center of the satellite between the Saturn-to-Titan line and the point in Titan where the potential is being evaluated, $R$ is the distance from the center of Titan to the evaluation point, and $r$ is the Saturn-to-Titan distance.

\par $\cos\alpha$ is given by

\begin{equation}
  \label{eq:cosalfa}
  \cos\alpha=\frac{\vec{o}.\vec{s}}{rR}
\end{equation}
where $\vec{o}$ and $\vec{s}$ are, respectively the orbital position of Saturn and the position of the surface element in cartesian coordinates whose origin is the center of Titan.

\par The cartesian coordinates of a surface element of the satellite in the body-fixed frame $(\vec{f_1},\vec{f_2},\vec{f_3})$ are given by

\begin{equation}
  \label{eq:szero}
  \vec{s_0}=R\times(\sin\theta\cos\lambda, \sin\theta\sin\lambda, \cos\theta),
\end{equation}
$\lambda$ being the planetocentric longitude, and $\theta$ the colatitude. Then $\vec{s}$ is obtained by 5 successive rotations, the same as Eq.\ref{equ:passage}, i.e.

\begin{equation}
  \label{eq:s}
  \vec{s}=R_3(-l)R_1(-J)R_3(-g)R_1(-K)R_3(-h)\vec{s_0}
\end{equation}
Since the goal is just to obtain the contribution of the amplitude of $J$, I used $K=0$, $h=0$, $l=-wt$, and $g=(n+w)t$. In fact, since $K$ is null, $g$ and $h$ are not defined but $g+h$ is. Since the node of Titan does not circulate, it disappears in averaging over the periods of the motion. Here a constant value for $J$, $J_0$, is considered. In fact, $J$ is not constant, so $J_0$ could be the mean value of $J$, i.e. $<J>$.

\par With these conventions, $\vec{o}=r(\cos f,-\sin f,0)$, $f$ being the true anomaly. 

\par For small eccentricity,

\begin{equation}
  \label{eq:recop1}
  r^{-1}=a^{-1}(1+e\cos nt)
\end{equation}

\begin{equation}
  \label{eq:recop2}
  \cos f = \cos nt +e(\cos 2nt-1)
\end{equation}

\begin{equation}
  \label{eq:recop3}
  \sin f = \sin nt + e \sin 2nt
\end{equation}
and finally $dS$ is given by 

\begin{equation}
  \label{eq:dS}
dS=\sin\theta d\theta d\lambda.
\end{equation}
\par For $J_0=0$ and a nonzero eccentricity one obtains (see Wisdom 2004 \cite{Wisdom2004}):

\begin{equation}
  \label{eq:classicaldE}
  \frac{dE}{dt}=-\frac{21}{2}\frac{3h_s}{5}\frac{GM_{\saturn}^2R_T^5ne^2}{a^6}\Delta
\end{equation}

\par With $e=0$ and $J_0$ nonzero, I obtain:

\begin{equation}
  \label{eq:moidE2}
  \frac{dE}{dt}=-\frac{3}{2}\frac{3h_s}{5}\frac{GM_{\saturn}^2R_T^5J_0^2(n+w)^2}{na^6}\Delta.
\end{equation}
Assuming $k_s=3h_s/5$ and replacing $\Delta$ by $-1/Q_s$ as in Wisdom \cite{Wisdom2004}, I get

\begin{equation}
  \label{eq:moidE3}
  \frac{dE}{dt}=\frac{3}{2}\frac{k_s}{Q_s}\frac{GM_{\saturn}^2R_T^5J_0^2(n+w)^2}{na^6}
\end{equation}
and finally:

\begin{equation}
  \label{eq:moidE5}
  \frac{dE}{dt}=\Big[\frac{21}{2}e^2+\frac{3}{2}(\sin I)^2+\frac{3}{2}J_0^2\Big(\frac{n+w}{n}\Big)^2\Big]\times\frac{k_s}{Q_s}f\frac{GM_{\saturn}^2nR_T^5}{a^6}
\end{equation}
where $f>1$ is an enhancement factor that takes a partially molten interior into account, and $I$ the obliquity of Titan (very small).

\par It is widely assumed that the tidal dissipation inside a synchronous satellite depends only on the orbital eccentricity, i.e. the other contributions are assumed to be negligible. The ratio $\kappa$ between the contribution of the eccentricity and the contribution of the wobble is:

\begin{equation}
  \label{eq:kappa}
  \kappa=\frac{1}{7}\Big(\frac{J_0}{e}\Big)^2\Big(\frac{n+w}{n}\Big)^2 \approx \frac{1}{7}\Big(\frac{J_0}{e}\Big)^2
\end{equation}
With $e=0.0289$ (i.e. TASS1.6 value \cite{VD1995}), $\kappa=1$ (i.e. the
contribution of the wobble in the tidal dissipation is as high as the contribution of the
eccentricity) for $J_0=4.381^{\circ}$). For $\kappa=0.01$, we have
$J_0=0.438^{\circ}$, so the contribution of the wobble in the dissipation is negligible under this
value. In addition to this tidal dissipation, the wobble induces Coriolis forces that provoke
stress and strains, responsible for a non-tidal dissipation. This last contribution is very small
and difficult to determine, because of its dependence on the internal structure of the satellite. The reader can find further explanation for instance in Efroimsky and Lazarian (2000 \cite{EL2000}).

\par We note that the expression of the contribution of the wobble in the dissipation depends on the frequency $n+w$ instead of $n$. This can be physically explained by the fact that the wobble is bound only to the planet (i.e. is not linked, for instance, with Titan's orbital plane), so the wobble added to the spin can appear as a sum of two motions, whose frequency is $n+w$. Since the spin is associated with a period of $15.94545$ days (TASS1.6 value \cite{VD1995}) and the wobble with a period of about $350$ years, the composition of these two motions corresponds to a period of $15.94346$ days. The implication here is that the observed rotation may be a little faster than that required for spin-orbit synchronisation. Richardson et al. (2004 \cite{RLM2004}) used the period of $15.945421\pm0.000005$ days (according to De Pater and Lissauer 2001 \cite{PL2001}) and detected a spin period of $15.9458\pm0.0016$ days. With a period of wobble of $350$ years and this period of spin, the detected period might be $15.94343$ days, which is consistent with their measurements at the three sigma uncertainty level.

\subsection{Application to the other satellites}

\par I used the analytical model presented in this paper to compute the equilibrium obliquity and the periods of the free librations of the Galilean satellites of Jupiter and of the Saturnian satellite Rhea (see Tab. \ref{tab:frekgalrhea}), using the dynamical and gravitational parameters listed in Tab.\ref{tab:dynothsat}.

\begin{landscape}
\begin{table}[ht]
  \centering
  \caption{Dynamical and gravitational parameters associated with the four Galilean satellites of Jupiter and the Saturnian satellite Rhea. The mean motion $n$, $e_1$, $\Gamma_0$, $\Gamma_1$ and $\dot{\ascnode}$ come from L1-1 ephemeris (Lainey et al. 2006 \cite{Lainey2006}) and TASS 1.6 (Vienne and Duriez 1995 \cite{VD1995}), the radii $R_s$ come from the IAU 2000 recommendations (Seidelmann et al. 2002 \cite{SAB2002}) that are the same as IAU 2006 (Seidelmann et al. 2007 \cite{SAA2007}). Nevertheless, here I have used the value of Thomas et al. (2006 \cite{THV2006}) for Rhea's radius, because it is consistent with the parameters of its gravity field derived by Iess et. al (2007 \cite{Iess2007}). The references of the other parameters are indicated in the Table. For Jupiter, I use $J_2=1.4736\times10^{-2}$ (Campbell and Synnott 1985 \cite{CS1985}) and $R_{\jupiter}=71492km$ (Seidelmann et al. 2002 \cite{SAB2002}). For Saturn, these parameters are given in Tab.\ref{tab:data}.}
  \begin{tabular}{l|ccccc}
     & J-1 Io & J-2 Europa & J-3 Ganymede & J-4 Callisto & S-5 Rhea \\
\hline
$n$ $(y^{-1})$ & $1297.2044714$ & $646.24512024$ & $320.76544424$ & $137.51159676$ & $508.00931975$ \\
$e_1$ & $4.15108\times10^{-3}$ & $9.35891\times10^{-3}$ & $1.42898\times10^{-3}$ & $7.37558\times10^{-3}$ & $9.713\times10^{-4}$ \\
$\Gamma_0$ & $3.14217\times10^{-4}$ & $4.04049\times10^{-3}$ & $1.59327\times10^{-3}$ & $3.8423\times10^{-3}$ & $2.9705\times10^{-3}$ \\
$\Gamma_1$ & $9.04178\times10^{-5}$ & $2.20043\times10^{-4}$ & $8.53478\times10^{-4}$ & $2.24539\times10^{-3}$ & $4.207\times10^{-4}$ \\
$\dot{\ascnode}$ $(y^{-1})$ & $-0.845589$ & $-0.207903$ & $-0.0456245$ & $0$ & $-0.17546762$ \\
$M_s$ & $4.70\times10^{-5} M_{\jupiter}$ \cite{AS1996}  & $2.56\times10^{-5} M_{\jupiter}$ \cite{AS1996} & $7.84\times10^{-5} M_{\jupiter}$ \cite{CS1985} & $5.60\times10^{-5} M_{\jupiter}$ \cite{CS1985} & $4.05841\times10^{-6} M_{\saturn}$ \cite{Iess2007} \\
$R_s$ (km) & $1821.46$ & $1562.09$ & $2632.345$ & $2409.3$ & $764.4$ \\
$J_2$ & $1.846\times10^{-3}$ \cite{AJ2001} & $4.355\times10^{-4}$ \cite{AS1998} & $1.268\times10^{-4}$ \cite{AL1996} & $3.110\times10^{-5}$ \cite{AJ1998} & $7.947\times10^{-4}$ \cite{Iess2007}  \\
$C_{22}$ & $5.537\times10^{-4}$ \cite{AJ2001} & $1.315\times10^{-4}$ \cite{AS1998} & $3.818\times10^{-5}$ \cite{AL1996} & $1.050\times10^{-5}$ \cite{AJ1998} &  $2.3526\times10^{-4}$ \cite{Iess2007}  \\
$\frac{C}{M_sR_s^2}$ & $0.377$ \cite{AJ2001} & $0.347$ \cite{AS1998} & $0.311$ \cite{AL1996} & $0.353$ \cite{AJ1998} & $0.372$ \cite{Iess2007} \\
  \end{tabular}
  \label{tab:dynothsat}
\end{table}
\end{landscape}

\begin{table}[ht]
  \centering
  \caption{Equilibrium obliquity and fundamental frequencies of the free librations for the Galilean satellites and Rhea, obtained with the model presented in this paper. The results are consistent with the ones given by Henrard for Io (2005 \cite{Henrard2005}) and Europa (2005 \cite{Henrard2005e}).}
  \begin{tabular}{l|ccccc}
Satellites & $K^*$ (arcsec) & $T_u$ & $T_v$ & $T_w$ \\
\hline
J-1 Io & $140.07$ & $13.31$ d & $159.16$ d & $225.17$ d \\
J-2 Europa & $1866.29$ & $52.57$ d & $3.59$ y & $4.84$ y \\
J-3 Ganymede & $824.02$ & $186.37$ d & $23.38$ y & $30.08$ y \\
J-4 Callisto & $1720.40$ & $2.44$ y & $209.32$ y & $356.54$ y \\
S-5 Rhea & $1320.98$ & $51.84$ d & $2.60$ y & $3.59$ y \\
  \end{tabular}
  \label{tab:frekgalrhea}
\end{table}

\begin{table}[ht]
\centering
\caption{Comparison of the periods of the free librations of Io and Europa given by different models. The results labelled "Henrard" come from (Henrard 2005 \cite{Henrard2005}) for Io and (Henrard 2005 \cite{Henrard2005e}) for Europa, the ones labelled "SONYR" come from (Rambaux \& Henrard 2005 \cite{RH2005}), and the ones labelled ``circular model'' come from Paper I.}
\begin{tabular}{l|cccc}
 & Henrard & SONYR & circular model & this paper \\
\hline
Io & & & & \\
$K^*$ (arcsec) & $157$ & & & $140$ \\
u & $13.25$ days & $13.18$ days & $13.31$ days & $13.31$ days \\
v & $159.39$ days & $157.66$ days & $160.20$ days & $159.16$ days \\
w & $229.85$ days & & $228.53$ days & $225.47$ days \\
\hline
Europa & & & & \\
$K^*$ (arcsec) & $1864.62$ & & & $1866.29$ \\
u & $52.70$ days & $55.39$ days & $52.98$ days & $52.57$ days \\
v & $3.60$ years & $4.01$ years & $3.65$ years & $3.59$ years \\
w & $4.84$ years & & $4.86$ years & $4.84$ years \\
\hline
\end{tabular}
\label{tab:comprambh}
\end{table}

\par Tab.\ref{tab:comprambh} gives a comparison with previous studies for Io and Europa. We can see a good agreement, except perhaps with the ``obliquity'' of Io. I think that the difference of about $10.8\%$ can be explained by the fact that Lainey's ephemeris have been updated between Henrard's study and this paper. In particular, Henrard used $\Gamma_0=3.60\times10^{-4}$ rad as given in Lainey's PhD thesis (2002 \cite{Lainey2002}), which is about $12\%$ higher than the value used in this study.

\par These data have been computed assuming that the satellites are in hydrostatic equilibrium, especially for the values of $\frac{C}{M_sR_s^2}$. Since the values of the periods of libration depend on $\frac{C}{M_sR_s^2}$, observing these librations should help estimate this parameter and test the hypothesis of hydrostatic equilibrium.

\par Unfortunately, the free librations should have been damped by dissipative effects, unless the amplitude associated is being forced by a resonance (as could be the case for Titan). But no resonance appears clearly in Tab.\ref{tab:frekgalrhea} (perhaps except between Ganymede's wobble $W$ and the node of Europa, with a regression period of 30.22 years).

\section{Conclusion}

\par The general analytical model given in this paper permits a first 3-dimensional description of the rotation of every synchronous body. More particularly, it gives the equilibrium position, the three fundamental frequencies of the free librations, and the main terms of the free solution. It is applied successfully to Titan, Io and Europa, in the sense that the results are consistent with the previous studies.

\par Moreover, this paper presents a study of a possible resonance involving Titan's wobble. With it, I show that if the capture into resonance occured, then the wobble could currently have a forced amplitude of several degrees, and so induce a significant internal tidal dissipation. Moreover, the forcing of the wobble could falsify the space-based detection of Titan's period of rotation to appear a little faster than it actually is, in particular because Titan's rotation axis would be significantly different from its pole axis of figure. Cassini results on Titan's rotation should give some information about that.

\par Even if the system is not in resonance, a quasi-resonant state may have a detectable effect on the free libration w.

\begin{acknowledgements}

\par I am indebted to several colleagues for fruitful discussions, especially Nicolas Rambaux, Anne Lema\^itre, Jacques Henrard, Alain Vienne and Michael Efroimsky. I also warmly thank the two referees, A. Dobrovolskis and an anonymous one, who pointed out a mistake in a previous version of this paper. This work benefited from the financial support of a FUNDP postdoctoral research grant.

\end{acknowledgements}

\begin{appendix}
  \section{Analytical expression of the resonant Hamiltonian}
\label{app:hamiltonres}

\par I here detail the Hamiltonian given by (Eq.\ref{eq:T2}), i.e.

\begin{displaymath}
  \mathcal{T}=\psi \Theta+\mu \Theta^2+\epsilon\sqrt{2\Theta}\cos\theta
\end{displaymath}

\par The coefficients $\psi$, $\mu$ and $\epsilon$ are respectively:

\begin{equation}
  \label{eq:psi}
  \psi=\omega_{\theta}+\dot{\Phi}_6-\dot{\phi}_6
\end{equation}

\begin{equation}
  \label{eq:mu}
  \begin{split}
    \mu=-\frac{n(1+\delta_s)}{64P^{*4}W^{*2}}\bigg(\delta_1\Big(120R^{*2}\Gamma_0^2-12R^{*2}W^{*2}+48\Gamma_0^2P^{*2}W^{*2} \\
+96\Gamma_0^2R^{*2}W^{*2}-228R^*P^*\Gamma_0^2+48W^{*4}R^*P^*-21W^{*4}R^{*2} \\
+96W^{*4}\Gamma_0^2P^{*2}+168W^{*4}\Gamma_0^2R^{*2}-348W^{*4}\Gamma_0^2R^*P^*+24R^*P^* \\
-192R^*\Gamma_0^2P^*W^{*2}-15R^{*2}-24W^{*4}P^{*2}+48\Gamma_0^2P^{*2}+24R^*P^*W^{*2}-8P^{*2}W^{*2} \\
+\big(2\frac{R^*}{P^*}-\big(\frac{R^*}{P^*}\big)^2\big)\big(-96\Gamma_0W^{*4}P^{*2}+84R^*\Gamma_0W^{*4}P^* \\
-48\Gamma_0P^{*2}+48\Gamma_0P^*R^*W^{*2}+60R^*P^*\Gamma_0-48\Gamma_0P^{*2}W^{*2}\big)\Big) \\
+\delta_2\Big(-120R^{*2}\Gamma_0^2+2R^{*2}W^{*2}-16\Gamma_0^2P^{*2}W^{*2}-16\Gamma_0^2R^{*2}W^{*2} \\
+228R^*P^*\Gamma_0^2+48R^*P^*W^{*4}-21R^{*2}W^{*4}+96P^{*2}W^{*4}\Gamma_0^2+168R^{*2}W^{*4}\Gamma_0^2-348R^*P^*W^{*4}\Gamma_0^2 \\
-24R^*P^*+40R^*P^*\Gamma_0^2W^{*2}+15R^{*2}-24W^{*4}P^{*2}-48\Gamma_0^2P^{*2}-8R^*P^*W^{*2}+8P^{*2}W^{*2} \\
+\big(2\frac{R^*}{P^*}-\big(\frac{R^*}{P^*}\big)^2\big)\big(-96\Gamma_0W^{*4}P^{*2}+84R^*\Gamma_0W^{*4}P^* \\
+48\Gamma_0P^{*2}-8R^*\Gamma_0P^*W^{*2}-60R^*P^*\Gamma_0+16\Gamma_0P^{*2}W^{*2}\big)\Big)\bigg)
  \end{split}
\end{equation}
and

\begin{equation}
  \label{eq:epsilon}
  \begin{split}
  \epsilon=-\frac{ne_1\Gamma_1}{4\sqrt{W^*}P^{*\frac{5}{2}}}(1+\delta_s)\Big(\delta_1\big(8P^{*2}-46R^*P^*W^* \\
+4R^{*2}+16P^{*2}W^*-14R^*P^*+20R^{*2}W^*+ \\
+\big(P^*\sqrt{R^*(2P^*-R^*)}\big)\big(21\Gamma_0+69\Gamma_0W^*-60W^*\Gamma_0\frac{R^*}{P^*}-12\Gamma_0\frac{R^*}{P^*}\big)\big) \\
+\delta_2\big(-8P^{*2}-46R^*P^*W^* \\
-4R^{*2}+16P^{*2}W^*+14R^*P^*+20R^{*2}W^*+ \\
+\big(P^*\sqrt{R^*(2P^*-R^*)}\big)\big(-21\Gamma_0+69\Gamma_0W^*-60W^*\Gamma_0\frac{R^*}{P^*}+12\Gamma_0\frac{R^*}{P^*}\big)\big)\Big)
  \end{split}
\end{equation}

\par We can see that $\epsilon$ contains $e_1$ and $\Gamma_1$ in its prefactor, which confirms the fact that there is no resonance when one of the two dynamical parameters is null.

\end{appendix}


\end{document}